# On the magnetic structures of 1:1:1 stoichiometric topological phases *Ln*SbTe (*Ln* = Pr, Nd, Dy and Er)

Igor Plokhikh[1,*], Vladimir Pomjakushin[2], Dariusz Jakub Gawryluk[1], Oksana Zaharko[2], Ekaterina Pomjakushina[1]

[1]Laboratory for Multiscale Materials Experiments, Paul Scherrer Institut, PSI, Villigen, CH-5232, Switzerland

[2]Laboratory for Neutron Scattering and Imaging (LNS), Paul Scherrer Institut, PSI, Villigen, CH-5232, Switzerland

*Corresponding author: igor.plokhikh@psi.ch; igor.plohih@gmail.com

**Abstract**

*Ln*SbTe (*Ln* – lanthanide) group of materials, belonging to ZrSiS/PbFCl (P4/*nmm*) structure type, is a platform to study the phenomena originating from the interplay between the electronic correlations, magnetism, structural instabilities and topological electronic structure. Here we report a systematic study of magnetic properties and magnetic structures of *Ln*SbTe materials. The studied materials undergo antiferromagnetic ordering at $T_N$ = 2.1 K (*Ln* = Er), 6.7 K (*Ln* = Dy), 3.1 K (*Ln* = Nd). Neutron powder diffraction reveals ordering with $k_1$ = (½ + δ 0 0) in ErSbTe, $k_2$ = (½ 0 ¼) in NdSbTe. DySbTe features two propagation vectors $k_2$ and $k_4$ = (0 0 ½). No long-range magnetic order is observed in PrSbTe down to 1.8 K. We propose the most probable models of magnetic structures, discuss their symmetry and possible relation between the electronic structure and magnetic ordering.

1. Introduction

Topological materials are in the focus of the solid-state research within the last few decades as they offer multiple practically important properties ranging from large magnetoresistance, ultra-high charge carrier mobility of Dirac and Weyl fermions as well long-period non-collinear spin textures (skyrmions) [1 – 10]. The latter were proposed to be utilized in the next generation memory storage devices. Among different groups of materials, those which show both, topological features in reciprocal space (Dirac or Weyl semimetals) and topological magnetic textures in direct space are of a particular interest [11, 12]. One of such families is *Ln*SbTe (*Ln* – lanthanide) [13 – 15], which on top of that features charge ordering instabilities [16 – 19]. Their electronic structure is reminiscent of the isostructural ZrSiS with Dirac nodal lines close to the Fermi level [20, 21]. On the other hand, antiferromagnetic skyrmion phase was proposed in Te-doped GdSbTe (GdSb$_x$Te$_{2-x-\delta}$) [17]. Besides that, fractional magnetization plateaus were recently observed in DySbTe [22], TbSbTe [23] and in Te-doped CeSbTe (CeSb$_x$Te$_{2-x-\delta}$) [24], as a consequence of interference between the charge density wave and the magnetic order. Despite mounting interest to some of the *Ln*SbTe compounds, there are only fragmentary reports on their microscopic magnetism [25 – 27]. Our recent neutron powder diffraction studies of *Ln*SbTe (*Ln* = Ho and Tb) reveal coexistence or competition between several magnetically ordered phases in these materials [26]. In the case of TbSbTe, we show that magnetic symmetry arguments hint to a multi-*k* magnetic ordering in this material, which serves as a prerequisite for topological spin-texture. Given the intriguing magnetic behavior of the *Ln*SbTe materials, this group merits a more detailed microscopic study. In the current contribution, we present neutron powder diffraction study of four members, *Ln* = Pr, Nd, Dy and Er. Whereas synthesis, crystal structure and basic physical properties of the first three compounds were already described in the literature [22, 28, 29], ErSbTe is reported for the first time.

## 2. Experimental section

Powder samples of *Ln*SbTe (*Ln* = Pr, Nd and Dy) were prepared from stoichiometric mixtures of the corresponding elements (Pr 2N5 Alfa Aesar, Nd 3N Fisher Scientific AG, Dy 3N ChemPur, Te 6 N Alfa Aesar, Sb 5N Alfa Aesar) through annealing in evacuated silica ampules. The starting materials were thoroughly chopped in a He-filled glove box, pressed into pellets and annealed at 800°C (24h) and 900°C (12h) with intermediate grinding. In the case of ErSbTe, direct reaction of the elements (Er 3N ChemPur, Sb and Te) using the same temperature treatment yields samples that besides the target phase contain significant amount of ErTe and unreacted Sb. In this case, to suppress the formation of ErTe, an extra amount of Sb was added according to the stoichiometry Er/Sb/Te – 1/1.3/1. At the last step, the sample was cooled from 900°C to room temperature over 24 h, which led to precipitation of the over-stoichiometric Sb in form of drops. This procedure yields nearly phase pure samples of 1:1:1 ErSbTe, as confirmed through X-ray and neutron powder diffraction.

Phase compositions of the samples was checked and the data for structure refinement were collected at room temperature using a BRUKER AXS D8

ADVANCE diffractometer (CuKα radiation, Bragg-Brentano geometry, 1D LynxEye PSD detector, 5 – 120° 2θ range). DC magnetization of the samples was measured using a MPMS XL 7 T, Quantum Design magnetometer in 1.8 K – 300 K temperature range and 0 – 7 T field range. Sintered polycrystalline pieces loaded in gelatin capsules were used for magnetization measurements. Experimental crystallographic data (cell parameters and refined atomic coordinates) and magnetization as a function of field plots are provided in SI.

Neutron powder diffraction data has been collected on the HRPT diffractometer (SINQ PSI) [30] using the wavelength λ = 2.45 Å (Ge(400) monochromator) in 2θ range of 3.55–164.50°, and steps of 0.05°. The samples were loaded in a vanadium container, mounted in a sample changer into an Orange He-cryostat (1.5 – 310 K).

X-ray and neutron diffraction datasets were analyzed using the JANA2006/JANA2020 software using the standard mathematical approach for description of powder diffraction profile (Pseudo-Voight profile function, Legendre polynomials for background) [31, 32]. Search of *k*-vectors was done using the *k*-search code, implemented in the FullProf suite [33]. The representation and magnetic symmetry analysis [34] have been done using the ISODISTORT tool from the ISOTROPY software [35, 36] and some tools from the Bilbao crystallographic server (k-SUBGROUPSMAG, MAXMAGN) [37, 38]. The magnetic models for the determined propagation vectors were generated in the ISODISTORT, tested against the experimental diffraction patterns, ranked according to their profile residuals and the best candidates are described in text; summary of the results is provided in *Table1*. The magnetic structures were plotted using the VESTA visualization tool [39].

## 3. Results and discussion

**Table 1.** Summary of the magnetic structure analysis for $Ln$SbTe ($Ln$ = Er, Nd, Dy and Pr): magnetic propagation vectors deduced from NPD, representations and magnetic space groups providing the best fits as well as the basis transformation from the original tetragonal ($P4/nmm$) cell.

| Compound | Magnetic vector | Representation | Magnetic space group | Basis |
|---|---|---|---|---|
| $Ln$ = Er | $k_1$ = (½ + δ 0 0), δ = 0.0039(6) | mDT1 | $Pmmn.1'(a00)000s$ (alternative setting of $Pmnm.1'(00g)000s$, No. 59.1.9.4.m406.2) | (1, 0,0 0, 1, 0) 0, 0,1) |
| $Ln$ = Dy | $k_4$ = (0 0 ½) | mZ5- | $Pnma.1'_a[Pnmm]$ (No. 62.450) | (0, 0, 2 1, 0, 0 0, 1, 0) |
| | $k_2$ = (½ 0 ¼) | mW1W3 | $P2_1/m.1'_a[P2_1/m]$ (No. 11.55) | (0, 2, 0 1, 0, 0 0, -1, -2) |
| | $k_2$ and $k_4$ | mZ5- + mW1W3 | $P2_1/c'$ (No. 14.78) | (0, 1, 2 1, 0, 0 0, -2, 0) |
| $Ln$ = Nd | $k_2$ = (½ 0 ¼) | mW2W4 | $Pc.1'_c[Pm]$ (No. 7.28) | (0, 1, 2 -1, 0, 0 0, -2, 0) |
| $Ln$ = Pr | no long-range order ||||

### 3.1. NdSbTe

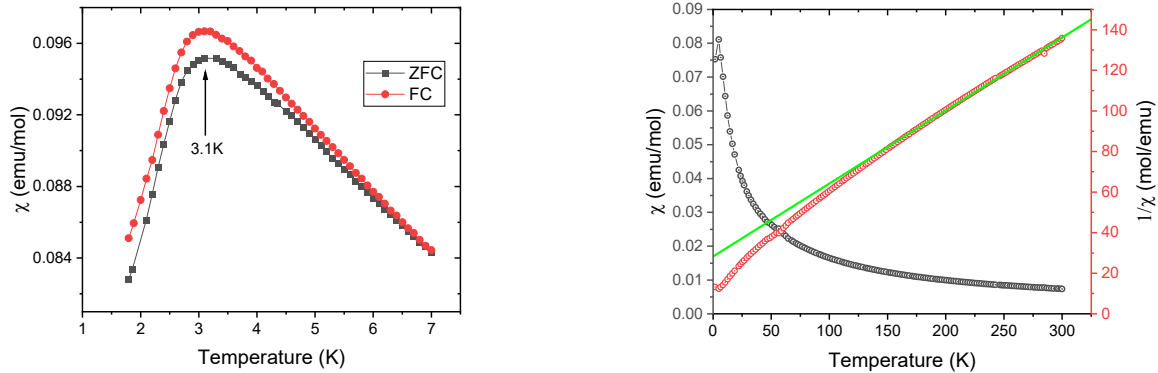

***Figure 1.*** (left) Field-cooling and zero-field cooling magnetic susceptibility curves for NdSbTe measured in 0.01 T field and (right) magnetic susceptibility in 1 T, inverse magnetic susceptibility, and the Curie–Weiss fit.

As follows from the magnetization measurements (***Figure 1***), our powder sample of NdSbTe undergoes a slightly broadened antiferromagnetic transition at $T_N$ =

3.1 K, which is consistent with the literature [29]. Above 150 K the sample behaves as a Curie-Weiss paramagnet with $\theta_{CW}$ = - 27.2(2) K and $\mu_{eff}$ = 3.90(2) $\mu_B$/Nd$^{3+}$. The value of magnetic moment is consistent with the Nd$^{3+}$ free ion value and the negative Curie-Weiss temperature reflects predominantly antiferromagnetic interactions.

The comparison of neutron diffraction patterns measured below and above the $T_N$ (**Figure 2**) shows multiple weak magnetic reflections in the antiferromagnetic state. All of them can be indexed using a single propagation vector $k_2$ = (½ 0 ¼). This is the W(½ 0 g) point of the Brillouin zone, with g locked to the commensurate value ¼. This point yields four complex irreps, which combine in pairs, giving two physical irreducible representations, mW1W3 and mW2W4 (see ref. 26). Magnetic models corresponding to the second representation fit the magnetic reflections; the corresponding group-subgroup tree is provided in supplementary information (SI). The *Pc*.1'$_c$[*Pm*] magnetic superspace group (MSSG) having four symmetrically independent magnetic Nd atoms and eight degrees of freedom (components of magnetic moment) allows us to construct a constant-moment magnetic structure. The magnetic moments of 1.7(1)$\mu_B$ are confined within the *a-c* plane of the initial cell and follow the pattern shown in **Figure 3**. Note, that the derived monoclinic model, although does not retain 90° angles for the unit cell, is just a four-fold superstructure of the parent tetragonal cell, *i.e.,* has smaller volume than the one with eight-fold superstructure anticipated from the propagation vector.

This spin structure is reminiscent to that reported recently for NdSb$_{0.94}$Te$_{0.92}$ [40]. The magnetic order there is associated with the propagation vector (0 ½ ½) with magnetic reflections being broadened. Here we show however, that the stoichiometric and ordered NdSbTe shows sharp (resolution limited) reflections. Hence, the origin of the magnetic reflections broadening in NdSb$_{0.94}$Te$_{0.92}$ can be due to defects in the structure.

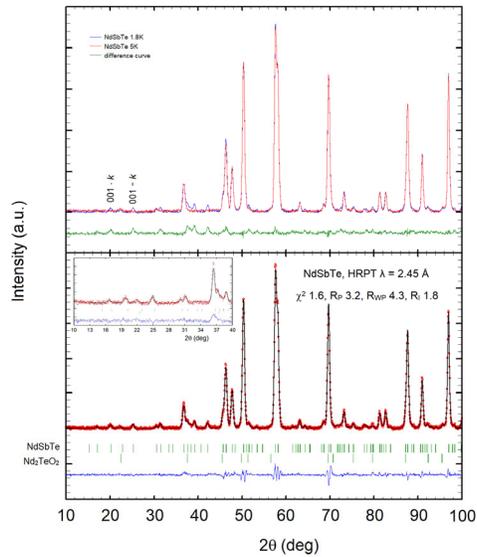

*Figure 2.* (top) Comparison of 1.8 K and 5 K NPD data for NdSbTe. The difference curve reveals the magnetic contribution. (bottom) Rietveld refinement plot for 1.8 K data. Small impurity (0.65(8)%) of $Nd_2TeO_2$ is included in the refinement. Enlarged low-angle region is shown on the insert.

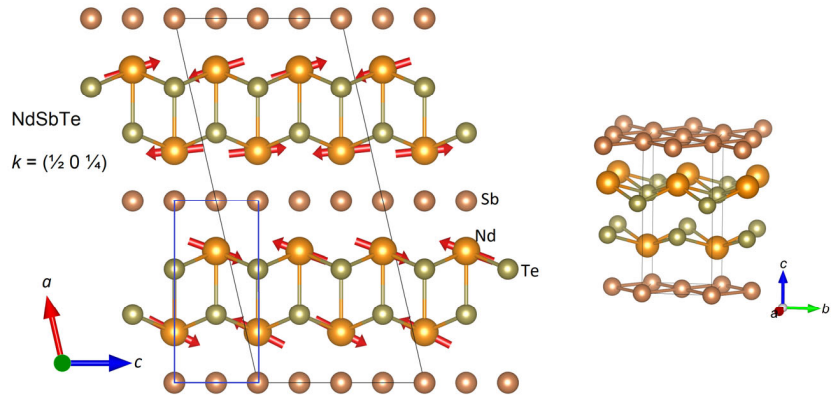

*Figure 3.* (left) Magnetic structure of NdSbTe. The magnetic unit cell is outlined in black, the parent nuclear – in blue. The coordinate system is given relative to the magnetic unit cell. (right) schematic representation of the parent nuclear structure.

## 3.2. PrSbTe

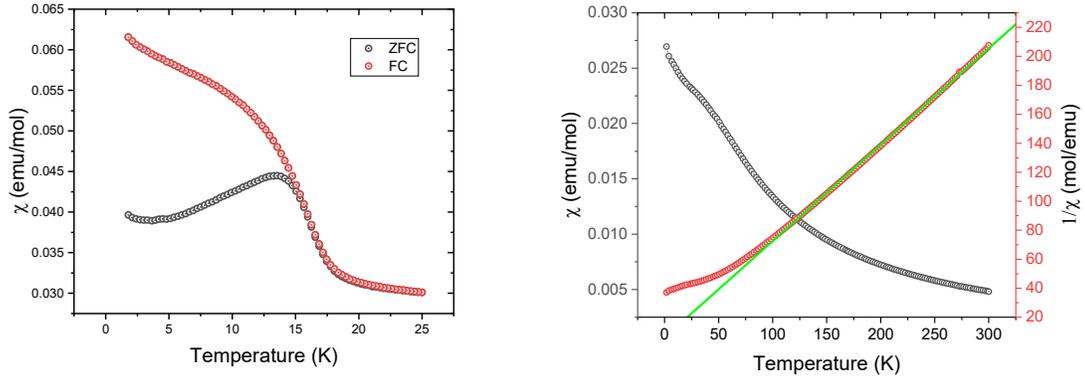

*Figure 4.* (left) The magnetic susceptibility vs. temperature curve measured in 0.01 T field for PrSbTe and (right) magnetic susceptibility in 1 T, inverse magnetic susceptibility, and the Curie–Weiss fit.

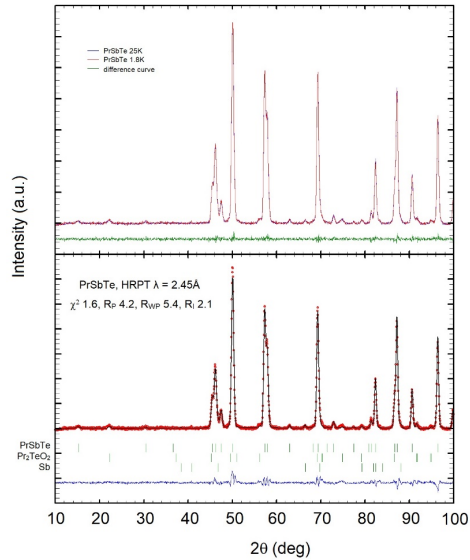

*Figure 5.* (top) comparison of 1.8 K and 25 K (blue) neutron diffraction data as well as their difference (green) for PrSbTe. (bottom) Rietveld refinement plot for 1.8 K data. Small impurities of $Pr_2TeO_2$ (0.9(1)%) and Sb (2.2(1)%) were included in the refinement.

PrSbTe behaves like a Curie-Weiss paramagnet above 100 K with $\theta_{CW}$ = - 29.7(2) K and $\mu_{eff}$ = 3.92(2) $\mu_B$/$Pr^{3+}$ (*Figure 4*). Around 18 K, the sample shows a transition-like feature, which is reminiscent of a ferromagnetic transition. On the other hand, the Curie-Weiss temperature is negative, indicative of predominantly antiferromagnetic interactions. Although there are no detailed studies in literature on PrSbTe, a short report available as Ref. 41 indicates, that single-crystals of this compound show no transitions down to 1.8 K, which might serve as an indication, that the transition-like feature observed in our magnetization measurements is extrinsic. Indeed, the neutron diffraction patterns measured at 1.8 K and 25 K are nearly identical as follows from *Figure 5*; they both can be described by the nuclear model of the main phase and small impurities, without magnetic Bragg scattering. Absence of magnetic ordering is likely due to the singlet crystal-electric field (CEF) configuration, which also precludes magnetic ordering in, for instance, metallic Pr (see explanation in ref. [42]).

### 3.3. DySbTe

There is a single study of bulk magnetic properties of DySbTe, where the authors show that this compound in a single-crystalline form orders antiferromagnetically at 7 K, [22] as deduced from magnetization, specific heat and resistivity measurements. Our powder sample shows a sharp transition at 6.7 K as provided in *Figure 6*. A small deviation can be ascribed to polycrystalline effect (defects, minor deviations from stoichiometry). Above 50 K, the sample behaves as a Curie-Weiss paramagnet with $\mu_{eff}$ = 10.67(2) $\mu_B$/Dy$^{3+}$ and $\theta_{CW}$ = - 10.8(1) K. The value of the paramagnetic moment is in a good agreement with the theoretical value (10.6 $\mu_B$) and the negative Curie-Weiss temperature is consistent with the antiferromagnetic nature of the transition.

Similar to *Ln*SbTe (*Ln* = Ho and Tb), some of the additional reflections in the neutron powder diffraction pattern measured at the base temperature (*Figure 7*) can be described using $k_2$ = (½ 0 ¼) according to the notation in ref. 26. Besides that, there are other reflections, corresponding to $k_4$ = (0 0 ½). At low 2θ angles, the reflections from the both propagation vectors have intensities comparable to the main reflections, whereas they are nearly extinct at high angles; thus, they were judged to be of a magnetic origin. We note that $k_4$ = 2 · $k_2$, so the $k_4$ reflections are formally the second-order satellites of $k_2$. However, the 2$^{nd}$ order satellites are highly improbable to have the magnetic origin [34]. Below we consider two models: a phase separation and a double-*k* models, which fit the data equally well. On the one hand, by the analogy with HoSbTe and TbSbTe [26], we can tentatively propose, that $k_2$ and $k_4$ correspond to different spatially separated phases within the same crystalline phase. On the other hand, unlike for HoSbTe and TbSbTe, diffraction patterns measured within the magnetically ordered state, reveal no region where one of the vectors disappears, *i.e.* both persist in couple up to the transition temperature.

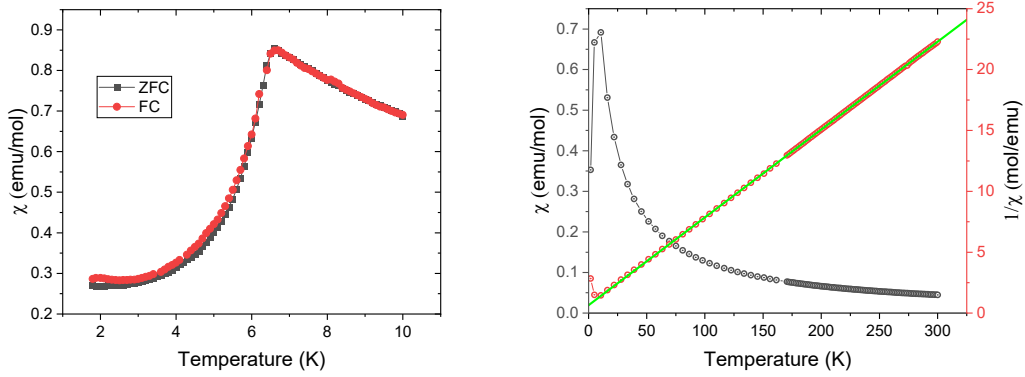

*Figure 6.* (left) Zero-field cooling and field-cooling magnetic susceptibility vs temperature curves for DySbTe measured in 0.01 T field and (right) magnetic susceptibility in 1 T, inverse magnetic susceptibility, and the Curie–Weiss fit.

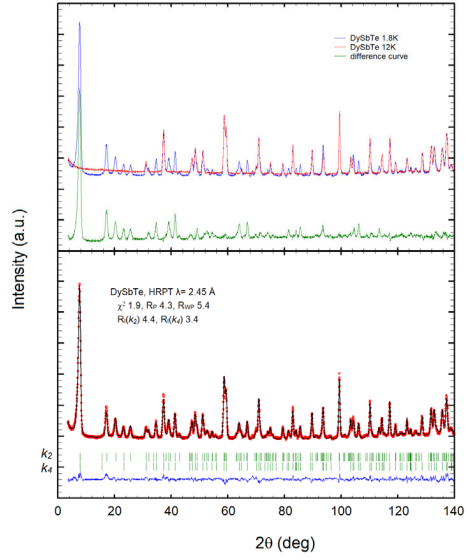

*Figure 7.* (top) Comparison of 1.8 K and 12 K neutron powder diffraction patterns. The difference between them reveals magnetic scattering due to the antiferromagnetic transition. (bottom) The Rietveld refinement plot for 1.8 K data.

The magnetic ordering vectors $k_4$ and $k_2$ correspond to mZ (0 0 ½) and mW (½ 0 g), g = ¼ points of the Brillouin zone respectively. Magnetic structures corresponding to the maximal magnetic space groups of mZ5- and mW1W3, $P2_1/m.1'_a[P2_1/m]$ and $Pnma.1'_a[Pnmm]$ respectively, fit well the magnetic intensities. Both solutions are collinear with the magnetic moments of the Dy atoms lying along the *a*-axis according to the pattern shown in *Figure 8*. Single-crystal magnetization measurements provided in ref. 22, show that the magnetic moments of the $Dy^{3+}$ ions lie within the *ab*-plane, which is consistent with our results. Double-*k* models were looked for among mixed mZ5- and mW1W3 representation, first considering only one arm of $k_2$ (½ 0 ¼), with additional requirement to have an equal magnetic moment on all Dy atoms. The magnetic model $P2_1/c'$ with two symmetrically independent Dy atoms provides the same quality of fit.

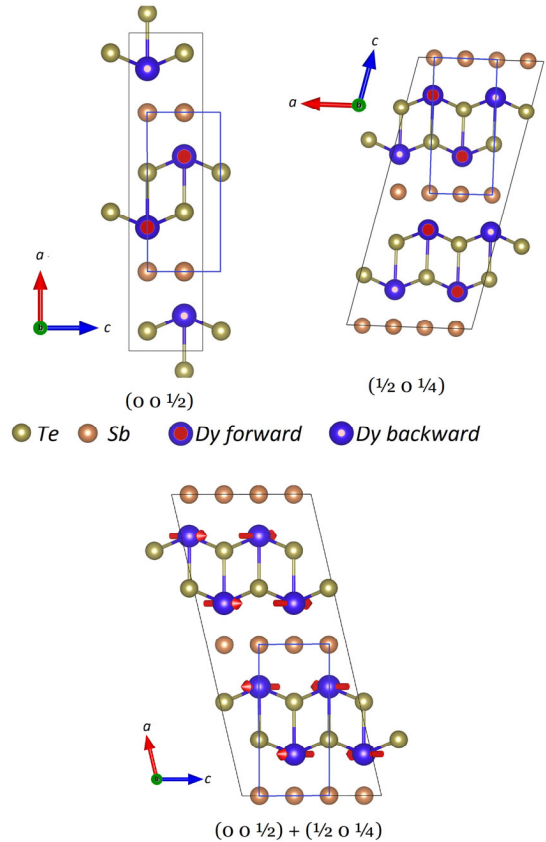

*Figure 8.* Magnetic structures of DySbTe corresponding to two-phase model (top) for $k_2$ (right top) and $k_4$ (left top) and combined double-*k* $k_2$ and $k_4$ (bottom) The magnetic unit cells are outlined in black, the parent nuclear – in blue. The coordinate systems are given relative to the *magnetic* unit cells.

Both two-phase and double-*k* models recover a significant part of the theoretical magnetic moment of $Dy^{3+}$ ($M_{theor} = g_J \cdot J = 10\ \mu_B$). The two-phase model (43.3(3)% of $k_2$ and 56.7(3)% of $k_4$) gives $M_{exp} = 7.43(1)\ \mu_B$., while the double-*k* model - $M_{exp} = 7.6(1)\ \mu_B$. Hence, they both are physical; a single-crystal neutron diffraction study using applied magnetic field can address this ambiguity. The presence of fractional plateaus in magnetization as a function of field data [22] hints at the two-phase model with these two phases responding differently to field, similar to the case of TbSbTe [23].

### 3.4. ErSbTe

ErSbTe has not been reported so far neither in the polycrystalline nor in the single-crystal form. As it is already mentioned in experimental section, our first attempts to prepare this material using the procedure, which was applied for Pr-, Nd-, Dy- samples, were not successful, due to the presence of a significant amount (*ca.* 20%) of the binary ErTe along with the main phase ErSbTe in polycrystalline batches. To suppress the formation of the binary phase, an excess of Sb was added to the reaction mixture. This yields the sample that is nearly free of ErTe; the amount of the latter in the sample used for NPD was estimated to be 1.0(1) %. The refinement of neutron powder diffraction data in the paramagnetic state (nuclear contribution only) indicates a good agreement with the 1:1:1 stoichiometry.

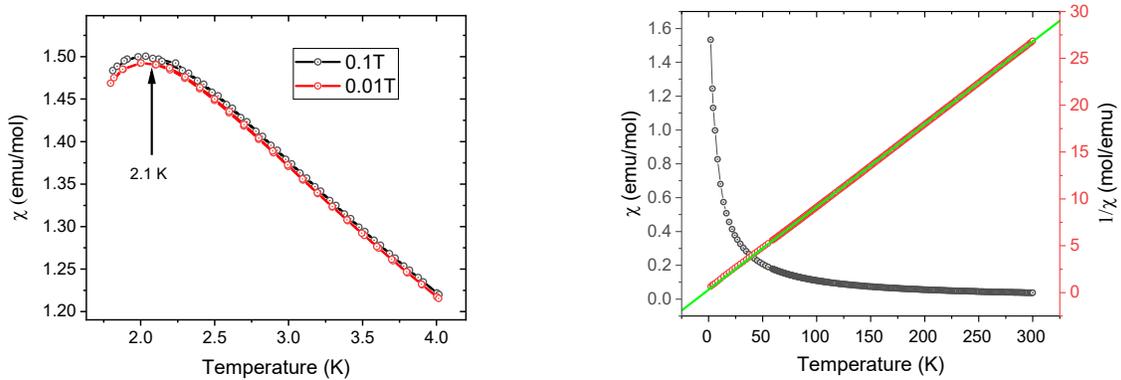

*Figure 9.* (left) Magnetic susceptibility *vs.* temperature for ErSbTe measured in 0.1 T and 0.01 T fields. The ZFC-FC curves measured in 0.01 T coincide. (right) Magnetic susceptibility in 1 T, inverse magnetic susceptibility, and the Curie–Weiss fit.

According to the magnetization measurements shown in *Figure 9*, the material remains paramagnetic down to very low temperatures. Only below 2.5 K the magnetization curve shows signatures of approaching the magnetic phase transition. The Néel temperature for this compound is *ca.* 2.1 K, *i.e.* very close to the base temperature of the MPMS. Above 50 K, magnetic susceptibility follows the Curie-Weiss law with $\theta_{CW}$ = - 6.07(6) K and $\mu_{eff}$ = 9.70(2) $\mu_B$/Er$^{3+}$. Whereas the magnetic moment is consistent with the Er$^{3+}$ free ion value of 9.6 $\mu_B$, the negative Curie-Weiss constant indicates predominantly antiferromagnetic interactions. The NPD pattern collected at 1.55 K besides the nuclear reflections features several strong magnetic ones (*Figure 10*), which confirms that the feature in magnetic susceptibility originates from the bulk magnetic ordering of the target material.

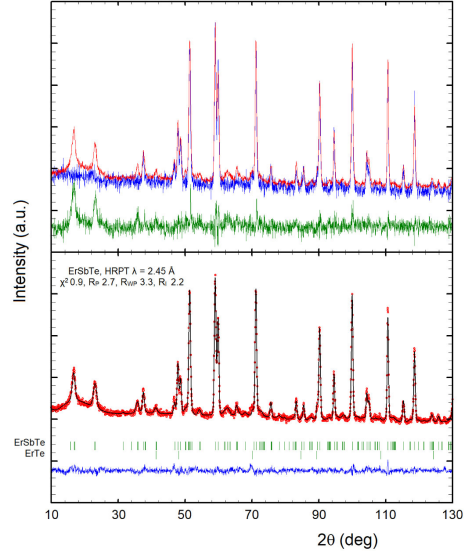

*Figure 10.* (top) Comparison of 1.55 K and 5 K NPD patterns as well as their difference revealing magnetic reflections. (bottom) Rietveld refinement plot for the magnetic structure at 1.55 K. The minor impurity of ErTe was included in the refinement.

The magnetic reflections in the 1.55 K NPD pattern could be indexed with the propagation vector very close to $k_1$ = (½ 0 0). Still, the Le-Bail fit (provided in SI) reveals a certain discrepancy. The magnetic reflections (½ 0 $l$, $l$ = 0, 1 and 2) are slightly broadened and displaced. This indicates a tiny deviation from commensurability. Indeed, $k_1$ = (½ + δ 0 0) with δ = 0.0039(6) combined with anisotropic peak broadening model [43] provides a better fit.

Testing the magnetic models corresponding to the mDT point results in mDT1 with as the maximal magnetic superspace group *Pmmn*.1'(*a*00)000*s* to give the adequate fit with a single variable component of the magnetic wave, $M_y\sin$ = 3.87(3) $\mu_B$, *i.e.* the modulation takes place in the direction perpendicular to the propagation vector. Any attempts to reduce the symmetry and to introduce other variables do not improve the fit and do not yield a constant-moment magnetic structure. This is a long period ($a/\delta \sim 100$ nm) transversal spin-density wave with the nearly antiferromagnetically coupled magnetic moments on the neighboring atoms. We should mention that since the NPD pattern was collected just slightly below the $T_N$, it might be that we probe not the ground-state structure. This is also evident from a pronounced diffuse scattering at $2\theta$ *ca.* 20deg. Measurements at lower temperatures might be necessary to address this issue.

## 4. Conclusions

In this study, we extend the knowledge on magnetic structures of the magnetic topological Dirac nodal-line semimetals belonging to the *Ln*SbTe family. We found that magnetic ordering in these compounds can be described using a restricted number of the commensurate (or nearly commensurate) propagation vectors: $k_1$ = (½ 0 0) in *Ln* = Tb, Ho and Er, $k_2$ = (½ 0 ¼) in *Ln* = Nd, Tb, Dy and Ho, $k_3$ = (½ 0 ½) in *Ln* = Tb and Ho, $k_4$ = (0 0 ½) in *Ln* = Ce and Dy. This similarity in magnetic orders might originate from similar nesting instabilities in the Fermi surface, which can be connected to the topological features of the electronic structure. The coexistence of several propagation vectors was also observed in *Ln* = Dy, besides those in *Ln* = Ho and Tb from the previous report. Collinear antiferromagnetic structures with the magnetic moments along the *a*-axis of the tetragonal cell were observed in DySbTe; the magnetic structure of NdSbTe is coplanar with the magnetic moments confined to the *ac*-plane; in ErSbTe, an incommensurate long period spin-density wave with the magnetic moments along the *a*-axis was detected.

## 5. Acknowledgements

The neutron powder diffraction experiments were performed at the Swiss spallation neutron source SINQ, Paul Scherrer Institute (Villigen, Switzerland). We thank the Swiss National Science foundation grants No. 200020-182536/1, 200021_188706 and R'equip Grant No. 461 206021_139082 and SNI Swiss Nanoscience Institute for the financial support.

We acknowledge Prof. Juan Manuel Perez-Mato for pointing out the issue with the correct usage of the unified magnetic space-group symbol [44, 45].